# Scaling up FluidFlower results for carbon dioxide storage in geological media

A. R. Kovscek[1,*], J. M. Nordbotten[2,4], and M.A. Fernø[3,4]

[1]*Energy Science & Engineering, Stanford University*

[2]*Department of Mathematics, University of Bergen*

[3]*Department of Physics and Technology University of Bergen*

[4]*Norwegian Research Center, Postboks 22 Nygårdgaten, 5838 Bergen*

## Abstract

The partial differential equations describing immiscible, but soluble, carbon dioxide ($CO_2$) displacement of brine are developed including local mass-transfer effects. Scaling relationships for characteristic time among laboratory and representative storage formation conditions are found upon assumption that free-phase $CO_2$ transport during injection is dominated by convection. The implication is that an hour in the FluidFlower (large-scale visual model) scales to hundreds of years of elapsed time in the storage formation. The scaling criteria permit extrapolation of the effects of changes in parameters and operating conditions. Interphase mass transfer allows $CO_2$ to saturate the brine phase and such mass transfer is a significant nonequilibrium phenomenon. Significant mixing of $CO_2$ dissolved into formation brine with original brine is found experimentally and is also predicted. The magnitude of onset time for downward migrating fingers containing $CO_2$ is typically only a fraction of the duration of $CO_2$ injection and in general agreement with theoretical analysis in the literature. Predictions for onset time of convective mixing at representative storage formation conditions likewise teach that the onset time for viscous fingering is significantly less than the duration of $CO_2$ injection in some cases. The implications of this observation include that mixing of $CO_2$ with brine and the subsequent settling due to gravity are relatively rapid and coincide with the period of active $CO_2$ injection.

## Introduction

The "FluidFlower" is an important new experimental tool for exploring coupled transport, physical, and nonequilibrium processes accompanying carbon dioxide ($CO_2$) injection into saline storage formations with complex geological bedding. One of the primary outcomes of experiments conducted in the FluidFlower is detailed data sets of the spatial evolution of injected $CO_2$ as a free phase as well as dissolved in brine. Such data is needed for the development and validation of predictive tools for storage to build confidence in modeling capabilities. Additionally, the FluidFlower is an important tool for generating interest in $CO_2$ storage and educating the casual observer about short and long-term storage mechanisms in the subsurface. Hence, another important outcome is the educational aspect of the visual results.

---

[*] email: kovscek@stanford.edu

Viewing the interplay of $CO_2$ convection with geological heterogeneity allows researchers to communicate to a wide community the mechanisms by which $CO_2$ may be stored long term as well as the types of geological features that promote secure storage. In this sense, the FluidFlower follows a long tradition among the flow in porous media community of scaled physical models to understand complex coupled processes, e.g., (Basu & Islam, 2009). Important aspects of such experiments include translation of experimental time into an equivalent time in the subsurface and, importantly, an understanding of how processes such as the rate of interphase mass transfer and convective mixing of $CO_2$-laden brine differs between the physical, laboratory model and the field.

The subsurface engineering community is rich with studies where scaling criteria have been developed to understand laboratory results in the context of field applications. For example, Lozada and Farouq Ali (Lozada & Ali, 1987) examine the displacement of heavy oil by immiscible carbon dioxide and the solubility of $CO_2$ in liquids to understand the role of different operating conditions on physical model results. Additionally, Basu and Islam (Basu & Islam, 2009) and Islam and Farouq Ali (Islam & Ali, 1990) present studies of scaling among laboratory and field for chemical enhanced oil recovery. In many cases, porous media properties and pressure differ significantly from the field (Kimber & Ali, 1989). The general consensus is that it is practically very difficult to scale all operative mechanisms in experiments with complex physical and chemical processes, but it is possible to estimate time scaling with a degree of certainty in convectively dominated systems as well as to understand the differences in scaling between laboratory and field for interphase mass transfer, diffusive transport within a phase, gravity, and so on.

In many studies, mass transfer effects among phases in numerical models of reservoirs and aquifers are neglected and local thermodynamic equilibrium is assumed, e.g., (Adenekan et al., 1993). There is a need, however, to quantify interphase mass transfer during geological storage because the phases are not initially in equilibrium (Erfani et al., 2022; Lindeberg & Wessel-Berg, 1997; Weir et al., 1995). The rate of dissolution of $CO_2$ into brine is controlled by diffusion, but, importantly, the resulting denser fluid may settle downward in the storage formation under the action of gravity (Ennis-King & Paterson, 2005; Kneafsey & Pruess, 2010; Riaz et al., 2006). Such convective mixing enhances solubility trapping of $CO_2$.

Table 1. Survey of papers that develop and use scaling criteria for subsurface fluid injection processes. An exhaustive listing of papers is out of scope of this work.

|  | Reference | Comment |
|---|---|---|
| **Oil Recovery** |  |  |
| Immiscible $CO_2$ injection to recover crude oil | (Lozada & Ali, 1987) | nonequilibrium mass transfer of $CO_2$ to liquid phase |
| Aqueous phase chemical injection to aid recovery | (Basu & Islam, 2009) | scaled advection diffusion, dispersion, and retention |
| Unsteady mass and heat transfer | (Kimber & Ali, 1989) | complete set of scaling groups for steam injection |
| Gravity override of low density injectant | (van Lookeren, 1983) | gravity override of injectant comparing lab and field |
| In situ combustion of crude oil | (Islam & Ali, 1992) | nonisothermal, reactive transport |
| **Contaminant Removal** |  |  |
| Cleanup of spilled hydrocarbons | (Sundaram & Islam, 1994) | removal of trapped organic phase using surfactant solutions |
| **Miscible Fingering** |  |  |
| Onset of gravity driven convection | (Riaz et al., 2006) | dense $CO_2$ laden brine fingering through unsaturated brine beneath gas cap |
| Convective mixing during $CO_2$ storage | (Ennis-King & Paterson, 2005) | inspectional and dimensional analysis of brine fingers |
| 2D and 3D simulation of convective mixing | (Pau et al., 2010) | at long time $CO_2$ mass flux reaches a stabilized rate |
| **Gravity Drainage** |  |  |
| three-phase gravity drainage | (Grattoni et al., 2001) | capillary and Bond numbers need to be combined to describe gravity drainage |
| gas-assisted gravity drainage | (Sharma & Rao, 2008) | scaled physical model experiments of gravity drainage |
| **Convective Miscible Mixing** |  |  |
| Convective mixing | (Hassanzadeh et al., 2007) | early, middle, and late time mixing of $CO_2$ laden brine beneath a gas cap |
| Hydrodynamic dispersion and convective mixing | (Erfani et al., 2022) | included hydrodynamic dispersion in the analysis of onset time for viscous mixing |

An exhaustive review of the dimensionless groups that describe scaling among laboratory and field processes is beyond the scope of this manuscript. Table 1, however, was constructed to communicate the breadth of physical and chemical mechanisms addressed by previous studies. That is, Table 1 presents the long tradition in scaling of laboratory results to field conditions. Much of the work summarized in Table 1 originates from oil recovery efforts due to the economic importance of crude oil. Note the efforts in thermal, chemical, and water-based recovery. Studies related to gas injection and associated gas solubility in reservoir fluids as well as chemical enhanced oil recovery are especially relevant to this manuscript (Islam & Ali, 1990; Lozada & Ali, 1987).

With the backdrop above, this manuscript is the first to present a methodology for and analysis of the scaling of time between physical processes in the FluidFlower and a geological formation during $CO_2$ injection. Time scaling emerges because $CO_2$ storage of this type is convectively dominated. Local thermodynamic equilibrium is not assumed because results from the FluidFlower have exhibited mass transfer effects at the gas/brine interface as well as convective mixing of $CO_2$-laden brine. In short, scaling of immiscible $CO_2$ injection into a saline aquifer with partial equilibrium between the gas and brine phases is our objective. We proceed with a description of the storage zone in the FluidFlower that is analyzed, the model and simplifications, model nondimensionalization, the scaling groups that emerge, analysis of miscible viscous fingers that contribute to convective mixing, and discussion. The differences among physical processes in the FluidFlower and the field are then explored via the scaling results. Discussion and conclusions complete the paper.

## FluidFlower Overview

The FluidFlower is packed with sands of varying grain size to create different geometries of porous media resembling geological strata including variation in permeability with depth, traps for buoyant fluids as well as both sealing and permeable faults. Figure 1 shows a heterogeneous subregion of the FluidFlower where buoyant free-phase $CO_2$ (orange-red color) is injected on the left and accumulates beneath the lightly colored sealing layer composed of fine-grained sand. The seal prevents gas entry up to about a gas column height of 0.2 m at pressure conditions near atmospheric whereas the maximum height of the gas zone is 0.1 m given the anticlinal geometry of the barrier layer and the open boundaries. This maximum height is illustrated in Fig. 1 (c) where free phase $CO_2$ spills upward around the left edge of the storage zone.

Another notable aspect of Fig. 1 is the dissolution of $CO_2$ into the brine underlying the region containing free-phase $CO_2$. This $CO_2$ laden brine takes on a deep red color that is nearly carmine. At late times as shown in Fig. 1(c), this dense $CO_2$-laden brine falls downward through less dense $CO_2$-free brine exhibiting miscible fingers.

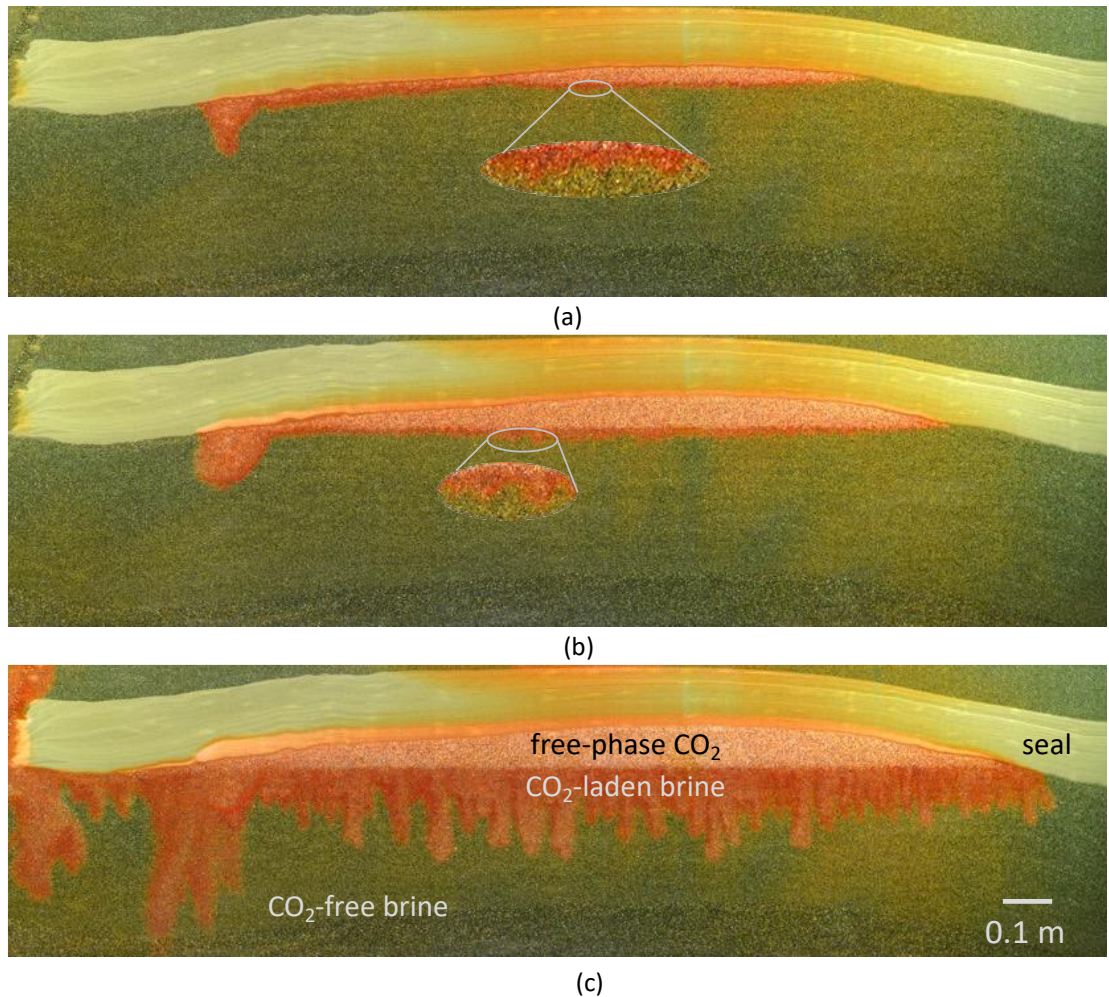

(a)

(b)

(c)

Figure 1. Representative images of filling of storage zone, saturation of underlying brine with dissolved $CO_2$, and viscous fingering of $CO_2$-laden brine into $CO_2$-free brine in the Fluidflower. In image (a) taken at 34 min post the start of $CO_2$ injection, diffusion/dispersion of $CO_2$ into the brine below the gas cap is evident with possible indications of initiation of viscous fingers as shown by inset image, (b) taken at 105 min displays expansion of the gas-filed zone and viscous fingers, while (c) taken at 647 min shows well-developed miscible fingers. The constant number of fingers in (b) and (c) suggests that little coalescence of fingers occurs over these time scales.

The FluidFlower is primarily a two-dimensional device because the depth is much less than the height and the width of the model. Differences between simulated two-dimensional and three-dimensional behavior of fingers revealed only modest differences in the time needed to form fingers as well as the downward flux of $CO_2$ mass (Pau et al., 2010). The third dimension did increase the complexity of fingers formed; however, the cumulative $CO_2$ mass flux was only about 25% greater for three-dimensional as compared to two-dimensional cases. This difference was viewed as small in comparison to the unknown and typically large variation in permeability subsurface that raises greater uncertainty in results. Hence, two-dimensional geometries are useful to understand storage formation dynamics. On the other hand, the storage formation geometry in Fig. 1 accentuates the vertical dimension somewhat. The ratio of the height to the width in the storage zone is about 0.05 in the FluidFlower whereas characteristic height and width from storage formations in Table 2 (Northern Lights and Sleipner) yield ratios of about 0.02. Hence, the vertical dimension is exaggerated by a factor of about 2 to 3.

## Model Description

This section presents a first-order model for processes in the FluidFlower. The analysis is limited to the storage zone shown in Fig. 1 and is two-dimensional. We progress from the main simplifications introduced to the dimensionless model to the scaling groups that emerge. It is assumed that the reader is acquainted with dimensional analysis and ordering, e.g., (Barenblatt, 1996; Denn, 1980)

**Main simplifications**

The following simplifications were made in the development of dimensionless equations and groups.
1. The variation of temperature is small across the system and so conditions are taken as isothermal.
2. There are two components denoted as $w$ and $c$ for water and carbon dioxide, respectively.
3. There are two phases in which the two components are mutually soluble. These phases are $b$ and $g$ denoting the brine-rich and $CO_2$-rich phases, respectively.
4. The multiphase extension of Darcy's law describes the convection, $u$, of a phase, $\beta$, as

$$u_\beta = -\frac{k k_{r\beta}}{\mu_\beta} \cdot (\nabla p_\beta - \varrho_\beta \mathbf{g}) \tag{1}$$

where k is the permeability, $k_{r\beta}$ is the relative permeability, $\mu_\beta$ is the viscosity, $p_\beta$ is the pressure, $\varrho_\beta$ is the phase mass density, and **g** is the acceleration of gravity.

5. The dispersive flux of a component, $J_i$, in partially-saturated porous media is described as (Ogata & Banks, 1961)

$$J_i = -\phi S_\beta \rho_\beta D_{i\beta} \cdot \nabla \chi_{i\beta} \tag{2}$$

where ϕ is porosity, $S_\beta$ is the phase saturation, $\rho_\beta$ is the phase molar density, $D_{i\beta}$ is the dispersion coefficient tensor, and $\chi_{i\beta}$ is the mole fraction of component i in phase β. Clearly, $S_\beta$ and $\chi_{i\beta}$ each sum to 1.

6. Mass transfer between phases is described by a two-film interface model.
7. There is no sorption of components to solids.
8. There are no chemical reactions.
9. Description of the mechanisms occurring in a vertical cross section is sufficient.

In view of the near atmospheric pressure in the FluidFlower and small gas-phase viscosity, assumption of an inviscid $CO_2$ phase and then proceeding to a material balance is appealing as an additional simplification. Accordingly, the magnitude of kinematic viscosity ($\nu = \mu/\varrho$) was evaluated for each phase because convective mass flux is inversely proportional to ν. At FluidFlower conditions, $\nu_g$ is $8 \times 10^{-6}$ m²/s whereas $\nu_b$ is $1 \times 10^{-6}$ m²/s. At conditions approximating a storage formation (2.6 $\times 10^7$ Pa and 366 K), $\nu_g$ and $\nu_b$ are $8 \times 10^{-8}$ and $3 \times 10^{-7}$ m²/s, respectively. The similar magnitudes of ν, the relatively modest differences in ν between phases, and the crossover in the phase with maximum ν as pressure and temperature increase suggest that, while the $CO_2$ phase is quite mobile, $CO_2$ viscosity is not negligible in comparison to brine.

**Convective flux**

Equation (1) is rewritten before proceeding to a mole balance on $CO_2$ and brine components. Below, convection is split according to flow driven by pressure gradient and by gravity. To proceed, we express the hydrostatic pressure, $P_\beta$, as

$$P_\beta = -\int_0^z \bar{\varrho}_\beta(z)\, dz \qquad (3)$$

where the integration is from the bottom "surface" of the storage zone. That is, the base of the storage zone that is the position of brine-gas interface, Fig. 1(c). The symbol $\bar{\varrho}_\beta$ is the mass density of an equilibrium fluid at that depth for brine or the density of gas phase at the total gas volume. Then, subtracting and adding $P_\beta$ to Darcy's law, Eq. (1), as well as evaluating the gradient of terms with gravity yields

$$u_\beta = -\frac{kk_{r\beta}}{\mu_\beta}\left(\nabla \cdot (p_\beta - P_\beta) - (\varrho_\beta - \bar{\varrho}_\beta(z))\mathbf{g}\right) \qquad (4)$$

Equation (4) expresses the phase flux as with reference to the deviation from hydrostatic conditions. Both pressure difference and density difference terms tend to zero as the system approaches equilibrium.

## Dimensionless mole balance

The nondimensionalized mass balance of component "i" incorporating multiphase transport of multicomponent fluids by convection and dispersion in the FluidFlower is obtained by summing over phases $b$ and $g$ and considering transport in the x and z direction as

$$\frac{\partial}{\partial t_D}\left[\phi_D \sum_{\beta=b,g}(S_{\beta D}\rho_{\beta D}\chi_{i\beta D})\right] = \left(\frac{H}{L}\right)^2 \left(\frac{(kk_{rg})_{RH}}{(kk_{rg})_{RV}}\right)\frac{\partial}{\partial x_D}\left[\sum_{\beta=b,g}\left(\rho_{\beta D}\chi_{i\beta D}\frac{(kk_{r\beta})_{DH}}{\mu_{\beta D}}\frac{\partial \Pi_{\beta D}}{\partial x_D}\right)\right]$$

$$+\frac{\partial}{\partial z_D}\left[\sum_{\beta=b,g}\left(\rho_{\beta D}\chi_{i\beta D}\frac{(kk_{r\beta})_{DV}}{\mu_{\beta D}}\frac{\partial \Pi_{\beta D}}{\partial z_D}\right)\right]$$

$$-\frac{\Delta\varrho_{gR}g_R H}{\Delta p_{gR}}\frac{\partial}{\partial z_D}\left[\sum_{\beta=b,g}\left(\rho_{\beta D}\chi_{i\beta D}\frac{(kk_{r\beta})_{DV}}{\mu_{\beta D}}(\varrho_{\beta D}-\bar{\varrho}_{\beta D}(z))\mathbf{g}_D\right)\right]$$

$$+\frac{t_R D_R}{L^2}\frac{\partial}{\partial x_D}\left[\sum_{\beta=b,g}\left(\phi_D S_{\beta D}\tau_{\beta D}D_{i\beta D}\frac{\partial \chi_{i\beta D}}{\partial x_D}\right)\right]$$

$$+\frac{t_R D_R}{H^2}\frac{\partial}{\partial z_D}\left[\sum_{\beta=b,g}\left(\phi_D S_{\beta D}\tau_{\beta D}D_{i\beta D}\frac{\partial \chi_{i\beta D}}{\partial z_D}\right)\right]$$

$$+q_{iD} \qquad (5)$$

where the z direction is aligned with the direction of gravity. The subscript D denotes a quantity that has been nondimensionalized, the subscript R marks characteristic quantities, $L$ is the characteristic horizontal length, $H$ is the characteristic vertical dimension, $D_{i\beta}$ is the dispersion coefficient of component i in phase β, $D_R$ is a representative magnitude of dispersion, $g_R$ is the magnitude of gravitational acceleration, $\mathbf{g}_D$ indicates the direction of gravity, and $q_{iD}$ is the nondimensionalized source/sink term for component i. The ratio $\left(\frac{(kk_{rg})_{RH}}{(kk_{rg})_{RV}}\right)$ expresses the anisotropy in the effective permeability of phase "g" as the effective permeability in the horizontal dimension upon the effective vertical permeability. The spatial variable z is nondimensionalized by H whereas $x_D$ is equal to z/L.

Convection due to pressure gradient and gravity are separated in Eq. (5) to make subsequent evaluation of the magnitude of these driving forces relative to each other more straightforward. Nondimensionalization of gradient terms is achieved via differences in pressure and density. The dimensionless phase potential is taken as $\Pi_{\beta D} = (p_\beta - P_\beta)/(p_{gR} - p_{g0})$ where $p_{gR}$ is the average stabilized pressure in the formation resulting from injection and $p_{g0}$ is the average initial pressure. The reference density difference is evaluated as $\Delta\varrho_{gR} = \varrho_g(p_{gR}) - \varrho_g(p_{g0})$ consistent with and $\Delta p_{gR} = p_{gR} - p_{g0}$. Hence, $\varrho_{\beta D} - \bar{\varrho}_{\beta D}(z)$ is equal to $(\varrho_\beta - \bar{\varrho}_\beta(z))/\Delta\varrho_{gR}$.

The dimensionless groups of $\frac{\Delta\varrho_{gR} g_R H}{\Delta p_{gR}}$, $\frac{t_R D_R}{L^2}$, and so on help us to understand the relative importance of convection driven by gravity and dispersive transport respectively. Interphase mass transfer does not appear in Eq. (5) because mass of species "i" lost by one phase is balanced exactly by the mass gained by the other phase.

The characteristic time, $t_R$, was obtained during nondimensionalization by making the coefficient on the expression for convection due to the pressure gradient in the vertical direction, that is the second term on the right of Eq. (5), to be of order 1. Hence, the characteristic time is

$$t_R = \frac{\phi_R S_{gR} \mu_{gR} H^2}{\Delta p_{gR} (kk_{rg})_{RV}} \qquad (6)$$

where $(kk_{rg})_R$ is a characteristic permeability to the gas. This choice of characteristic time makes dimensionless mass accumulation and z-direction convection to be of order 1 and, consequently, asserts that convection driven by the pressure gradient is the main transport mechanism during injection. Note also that the inverse of the ratio $\mu_{gR}H/(kk_{rg})_{RV}\Delta p_{gR}$ defines a characteristic vertical Darcy velocity. The characteristic source/sink term follows as

$$q_R = \frac{\rho_{gR}\chi_{cgR}\Delta p_{gR}(kk_{rg})_{RV}}{\mu_{gR}H^2} \qquad (7)$$

Equation (5) gives a fundamental constraint on the dynamics of both the FluidFlower as well as field-scale systems. On the other hand, it is important to note that Eq. (5) by itself does not provide a closed system, but must be complemented by a phase partitioning model, constitutive relations, boundary conditions, and so on. The phase partitioning will, in itself, introduce a characteristic time scale, as is discussed separately in a later section.

Table 2. Baseline reference values for comparing FluidFlower and storage formation time scales and physical processes.

|  | FluidFlower | Northern Lights | Sleipner (Utsira) | In Salah (Krechba) |
|---|---|---|---|---|
| $k_R$ (m$^2$) | 2.79x10$^{-9}$ | 2.0x10$^{-13}$ | 2.5x10$^{-12}$ | 1.0x10$^{-14}$ |
| $\phi_R$ | 0.40 | 0.25 | 0.37 | 0.16 |
| $k_{rg}^o$ | 0.11 | 0.3 | 0.3 | 0.3 |
| $S_g$ | 0.88 | 0.50 | 0.5 | 0.5 |
| $H$ (m) | 0.1 | 170 | 200 | 20 |
| $L$ (m) | 2.0 | 10,000 | 10,000 | 5000 |
| net to gross | 1.0 | 0.35 | 0.70 | 1 |
| $\Delta p_{gR}$ (Pa) | 1000 | 2.50x10$^6$ | 1.30x10$^6$ | 1.11x10$^6$ |
| $\mu_{gR}$ (Pa-s) | 1.50x10$^{-5}$ | 5.33x10$^{-5}$ | 3.98x10$^{-5}$ | 4.31x10$^{-5}$ |
| $k_V/k_H$ | 0.4 | 0.1 | 0.1 | 0.1 |
| $\Delta\varrho_{gR}$ (kg/m$^3$) | 0.0187 | 34.2 | 54.6 | 230. |
| $D_R$ (m$^2$/s) | 1.8x10$^{-8}$ | 7.0x10$^{-9}$ | 3.0x10$^{-9}$ | 7.0x10$^{-9}$ |

## Scaled Processes

In practice, it is very difficult to scale all processes between the laboratory and the field when (i) the coupled physical processes are complex and (ii) the geometry and permeability of the porous medium are heterogeneous (Lozada & Ali, 1987). The aim of this section is to establish the scaling of time between the FluidFlower and the storage formation for convectively dominated flows and, importantly, to estimate the differences in the relative magnitudes of transport driven by gravity and dispersion as well as mass transfer from the $CO_2$-rich phase to the brine-rich phase. This analysis applies to conditions during injection and before $CO_2$ spills, Fig. 1.

## Time scaling between lab and field

Equation (5) and the resulting characteristic time were developed with the notion that convection is the primary transport process during active $CO_2$ injection. The relation between elapsed time in the storage formation, $t^{Form}$, and that in the FluidFlower, $t^{Flow}$, is determined by the ratio of characteristic times as

$$t^{Form} = t^{Flow} \frac{t_R^{Form}}{t_R^{Flow}} \qquad (8)$$

The characteristic time for the FluidFlower is estimated using Eq. (6) and used to scale experimental results between cases with parameters approximating the Northern Lights (Marashi, 2021), Sleipner (Chadwick, 2013; Chadwick et al., 2012), and In Salah projects (Bissell et al., 2011; Ringrose et al., 2009).

Table 2 lists the data used to describe these field projects and the FluidFlower case shown in Fig. 1. Some settling of sand is evident during repeated tests in the FluidFlower. The porosity was corrected for observed sand settling from 0.44 to 0.40 and the permeability estimated using the Carmen-Kozeny equation to decrease from $4.26 \times 10^{-9}$ m$^2$ to $2.79 \times 10^{-9}$ m$^2$ (Lake et al., 2014). Additionally, storage formation heights were multiplied by the net-to-gross ratio.

Table 3. The first three row contain dimensionless coefficient magnitudes as identified in Eq. (5). The final row compares the baseline characteristic time scale of the FluidFlower and storage formation (given in field years per laboratory hour).

| | FluidFlower | Northern Lights | Sleipner (Utsira) | In Salah (Krechba) |
|---|---|---|---|---|
| $\left(\frac{H}{L}\right)^2 \left(\frac{(kk_{rg})_{RH}}{(kk_{rg})_{RV}}\right)$ | 6.2x10$^{-3}$ | 3.5x10$^{-4}$ | 2.0x10$^{-3}$ | 1.6x10$^{-4}$ |
| $\frac{\Delta \varrho_{gR} g_R H}{\Delta p_{gR}}$ | 1.8x10$^{-5}$ | 8.0x10$^{-3}$ | 5.8x10$^{-2}$ | 4.1x10$^{-3}$ |
| $\frac{t_R D_{gR}}{H^2}$ | 7.7x10$^{-7}$ | 3.1x10$^{-6}$ | 5.3x10$^{-7}$ | 7.2x10$^{-6}$ |
| $t^{Form}/t^{Flow}$ (years/hour) | | 420 | 390 | 110 |

With the values in Table 2, Eq. (8) teaches that 1 hour in the FluidFlower is representative of hundreds of years in the formation as summarized in Table 3. For Northern Lights and Utsira conditions, an hour in the FluidFlower scales to about 400 years whereas for In Salah conditions an hour scales to about a hundred years. The differences are primarily affected by formation thickness and permeability.

The time to fill the storage layer in the FluidFlower such that $CO_2$ spills out of the trap is roughly 4 hours (250 min) in some experiments, Fig. 1. With the values in Table 2 and Eq. (6), the experimental time to spill scaled to storage formation conditions is 100's to 1000's of years. Simulations of storage at Northern Lights (Johansen formation) injected $CO_2$ at a rate of 1.6 Mt/y and the storage formation, as simulated, had a capacity of 21,680 Mt (Marashi, 2021). The time to fill the storage formation at this rate is about 6,000 years. The long times required to fill a large-capacity storage formation are consistent with $t^{Form}/t^{Flow}$ in Table 3.

**Horizontal convection**

The magnitude of the coefficient $\left(\frac{H}{L}\right)^2 \left(\frac{(kk_{rg})_{RH}}{(kk_{rg})_{RV}}\right)$ on the first term on the right of Eq. (5) captures the relative importance of pressure gradient driven convection in the horizontal direction. With values in Table 2 for the FluidFlower and the Northern Lights project, this coefficient is 0.006 and 0.0003, respectively, indicating that the vertical direction dominates during pressure-driven convection. Results for other cases are found in Table 3. Note the importance of the characteristic horizontal dimension to results. Decreasing L from 10,000 to 1000 m for the Utsira case increases the coefficient from 0.002 to 0.2. The summary in Table 3 supports the importance of convection in the vertical direction.

**Gravity driven convection**

The importance of gravity as a force for driving convection in the FluidFlower and the field is understood by computing the coefficient $\frac{\Delta \varrho_{gR} g_R H}{\Delta p_{gR}}$ in front of the third term on the right of Eq. (5). The magnitude of the coefficient as compared to a value of 1 instructs about the relative importance of gravity. Likewise, the ratio of the coefficient is a measure of the difference in the importance of gravity between the storage formation and the FluidFlower. Values from Table 2 are used again and results for the 4 cases are in Table 3. The density difference in Table 2 corresponds to the values of pressure in the pressure difference.

The magnitude of $\frac{\Delta \varrho_{gR} g_R H}{\Delta p_{gR}}$ for the FluidFlower is 2 x$10^{-5}$. This value is indicative of the importance of convection by pressure gradient in the FluidFlower. The coefficient rises to values of 0.008 and 0.06 for conditions representative of Northern Lights and Utsira storage formations, respectively, indicating that the role of gravity relative to pressure is greater in the field as compared to the FluidFlower. Likewise, that ratio of the coefficients for gravity (FluidFlower:storage formation) is of order 0.001 indicating that the FluidFlower under represents the role of gravity with respect to the storage formation during injection. That is, the influence of gravity segregation on mass transport is greater in the field. Importantly, the values of $\frac{\Delta \varrho_{gR} g_R H}{\Delta p_{gR}}$ less than 1 are indicative that pressure gradient contributes significantly to convection during active injection.

**Dispersive transport**

The magnitude of the coefficient $\frac{t_R D_{gR}}{H^2}$ preceding the fifth term on the right of Eq. (5) teaches about the relative importance of dispersive transport of a component in the z-direction. This section specifically examines dispersive transport of $CO_2$ in the brine phase as it is relevant to understand the miscible fingers in Fig. 1(c). The characteristic times $t_R^{Form}$ and $t_R^{Flow}$ developed earlier for the FluidFlower and Northern Lights examples are used. The diffusivity of $CO_2$ in the brine phase is taken as 1.9 x$10^{-9}$ m$^2$/s at FluidFlower conditions (Tamimi et al., 1994) and 7.0 x$10^{-9}$ m$^2$/s at Northern Lights conditions (Cadogan et al., 2014).

The dispersion coefficient is obtained from a compilation of measurements by (Jha et al., 2011; Lake et al., 2014). Specifically, we use Fig. 11 of Jha et al. that plots the ratio of dispersion coefficient upon diffusion coefficient ($D_c/\mathcal{D}_c$) versus velocity that is made dimensionless by the ratio of particle diameter upon diffusion coefficient. The average interstitial velocity in the z direction of the gas/liquid interface in the FluidFlower is obtained from images, Fig. 1, and the $\phi_D$ and $S_{gD}$ in Table 2 as 1.9 x$10^{-5}$ m/s. The particle diameter of the storage zone in Fig. 1 is 1.77 mm such that $D_c/\mathcal{D}_c$ for $CO_2$ in the brine phase is found as 10 and the dispersivity is 1.9 x$10^{-8}$ m$^2$/s. Similarly, taking the interstitial velocity at Northern Lights conditions as 0.1 m/d (6.6x$10^{-6}$ m/s), the particle diameter as 32 µm, and the diffusivity above yields $D_c/\mathcal{D}_c$ of about 1. We assume that this ratio is roughly 1 for the other cases as well.

With these values we find that $\frac{t_R D_{gR}}{H^2}$ is equal to 8 x$10^{-7}$ in the FluidFlower and ranges from $10^{-7}$ to $10^{-6}$ in the various storage formations. The small values are consistent with macroscopic transport being driven primarily by convection. Values of $\frac{t_R D_{gR}}{L^2}$ are even smaller because L is at least on order of magnitude greater in all cases.

**Interphase mass transfer**

Diffusion, while not a contributor to transport over large distances, is quite important to driving mass transfer across the interface between phases. Note the carmine-red layer of $CO_2$-laden brine beneath the gas zone in Fig. 1(b). Subsequently, locally dense brine phase sinks vertically through the model. Interphase mass transfer of $CO_2$ does not appear in the overall mole balance for chemical species, Eq. (5), because mass lost by the $CO_2$-rich phase is equal to the mass gained by the brine phase.

To understand mass transfer rates within the zone occupied by $CO_2$ and brine, we apply the so-called two-film model for mass transfer resistance at the interface between phases to quantify the mass transfer rate (Lewis & Whitman, 1924). Appendix A shows that the flux of $CO_2$ across the interface between the $CO_2$-rich phase and brine phase, J$_{cgb}$, is written as

$$J_{cgb} a_i = K a_i (\chi_{ci} - \chi_{cb}) \quad (9)$$

where $K$ is an overall mass transfer coefficient, $a_i$ is the interfacial area, $\chi$ is again mole fraction, the subscript *ci* refers to the amount of $CO_2$ in the brine phase at interface conditions, and the subscript *cb* refers to $CO_2$ in the bulk brine phase.

(Martin et al., 1981) measured the mass transfer between $CO_2$ and liquid phases in porous media under immiscible conditions and present a correlation for mass transfer resistance. We use this correlation and modify it as suggested by (Lozada & Ali, 1987) and include the pressure drop. That is, we apply Darcy's law to describe mass flux as well as divide by porosity and liquid phase saturation to obtain the interstitial phase velocity, $v_w$. The expression for mass transfer coefficient is

$$K a_i = B \mathcal{D}_{cb} \left( \frac{\Delta p_{gR}}{H} \right) \left( \frac{k k_{rb}}{\phi S_b \mu_b} \frac{\Delta p_{gR}}{H} \right) \quad (10a)$$

or

$$K a_i = B \mathcal{D}_{cb} \left( \frac{\Delta p_{gR}}{H} \right) v_w \quad (10b)$$

where B is determined by experiment (Martin et al. report 0.011). The diffusivity of $CO_2$ in brine, $\mathcal{D}_{cb}$, is the molecular diffusivity of $CO_2$ in the liquid phase accounting for the pore-scale nature of mass transfer from g to b phases.

The way ahead is to compute the ratio of interphase mass transfer as given by Eq. (9) between the FluidFlower and the representative formations. We set $\chi_{cb}$ equal to 0 in Eq. (9) because the largest mass transfer rates are experienced where the amount of $CO_2$ dissolved in the brine is small. The equilibrium solubility of $CO_2$ in brine at the interface ($\chi_{ci}$) is computed as described by (Enick & Klara, 1990) using their correlations and the Krichevsky-Ilinskaya equation (Prausnitz et al., 1999). The solubility is found to be $\chi_{ci} = 0.021$ at Northern Lights conditions and $\chi_{ci} = 6.8 \times 10^{-4}$ at FluidFlower conditions. The prediction for FluidFlower conditions agrees well with data ($\chi_{ci} = 7 \times 10^{-4}$) in Lange's Handbook (Lange, 2017).

Table 4. Parameter values to compute the ratio of interphase mass transfer and the onset of miscible fingering. Additional parameters needed for computation are taken from Table 2.

|  | FluidFlower | Northern Lights | Sleipner (Utsira) | In Salah (Krechba) |
|---|---|---|---|---|
| $k_{rb}$ | 0.17 | 0.10 | 0.10 | 0.10 |
| $S_b$ | 0.12 | 0.50 | 0.5 | 0.5 |
| $\mu_b$ (Pa-s) | $1.0 \times 10^{-3}$ | $3.1 \times 10^{-4}$ | $6.4 \times 10^{-4}$ | $3.0 \times 10^{-4}$ |
| $\Delta \varrho_b$ (kg/m$^3$) | 3.5 | 13 | 10.5 | 10.5 |
| $D_{cb}$ (m$^2$/s) | $1.9 \times 10^{-9}$ | $7.0 \times 10^{-9}$ | $3.0 \times 10^{-9}$ | $7.0 \times 10^{-9}$ |
| $\chi_i$ | $6.8 \times 10^{-4}$ | 0.021 | 0.021 | 0.020 |
| $Ka_i/B$ (Pa m$^2$/s) | $1.9 \times 10^{-6}$ | $1.9 \times 10^{-9}$ | $1.8 \times 10^{-10}$ | $2.6 \times 10^{-8}$ |
| $Ja_i^{Flow}/Ja_i^{Form}$ |  | 31 | 330 | 3 |

Equation (10) is substituted into Eq. (9) and the ratio of mass transfer in the FluidFlower relative to the storage formation is found. In this way, the coefficient B does not need to be evaluated. Parameters for calculation are taken from Table 2 and supplemented by Table 4. Equation (10a) is used for the FluidFlower whereas Eq. (10b) is used for the storage formation. It is anticipated that nonzero mass transfer from the gas to the brine phase occurs at the advancing interface in the storage formation because elapsed time is much greater and hence Eq. (10b) is more applicable. The interstitial brine velocity for the storage formation conditions is set to $6.6 \times 10^{-6}$ m/s consistent with the earlier discussion of dispersion.

The ratio between FluidFlower and storage formation conditions is roughly a factor of 30 for Northern Lights, 300 for Sleipner conditions, and about 3 for In Salah. These ratios greater than 1 primarily result because the permeability of the sands in the FluidFlower are 4 orders of magnitude larger than the storage formation and the values of $H$ differ by at least 2 orders of magnitude. The calculations summarized in Table 4 indicate that interphase mass transfer is faster in the FluidFlower relative to the formation and motivate the exploration of fingering and convective mixing that follows.

## Fingering and Convective Mixing

An outcome of the mass transfer described by Eq. (9) is the formation of a layer of dense $CO_2$-laden brine just beneath the capillary transition zone in the FluidFlower in Fig. 1. This $CO_2$-laden brine clearly segregates downward convectively in Fig. 1(c). Ultimately, mixing convolves mass transfer of $CO_2$ to the gas-brine interface, diffusion of $CO_2$ away from the interface and into the bulk brine, and the convection of denser $CO_2$-laden brine downward in the formation through the formation of viscous fingers.

Figure 2 presents an overview of dissolution of $CO_2$ versus time into the brine phase as found in the FluidFlower. The dissolved mass is obtained by subtracting the mass of $CO_2$ in the free phase from the cumulative injected mass. Figure 2 also plots the diffusive limit as a dashed line computed using $\mathcal{D}_{cb}$ from Table 2. Note the inset that presents results at very short times. The thickness of a $CO_2$ saturated layer is of order $2(\mathcal{D}t)^{1/2}$ (Ennis-King & Paterson, 2005). For a $CO_2$ diffusivity in brine of 1.9 x$10^{-9}$ m$^2$/s and an elapsed time of 3600 s, the thickness of a $CO_2$ laden layer is about 5 mm. Figure 2 indicates, however, that only very short times are controlled by diffusion across the gas-brine interface. Moderate times illustrate dispersion supporting the inclusion of dispersion in Eq. (5).

Consistent with linear stability analysis of miscible displacement (Elenius & Johannsen, 2012), Fig. 1 illustrates early onset of instability whereas Fig. 2 shows that this initial unstable regime follows $t^{1/2}$ scaling until about 150 minutes. The fully unstable system then emerges, mixing evolves, the rate of mass transfer increases, and the scaling of mass in solution transitions to $t$ (Hassanzadeh et al., 2007).

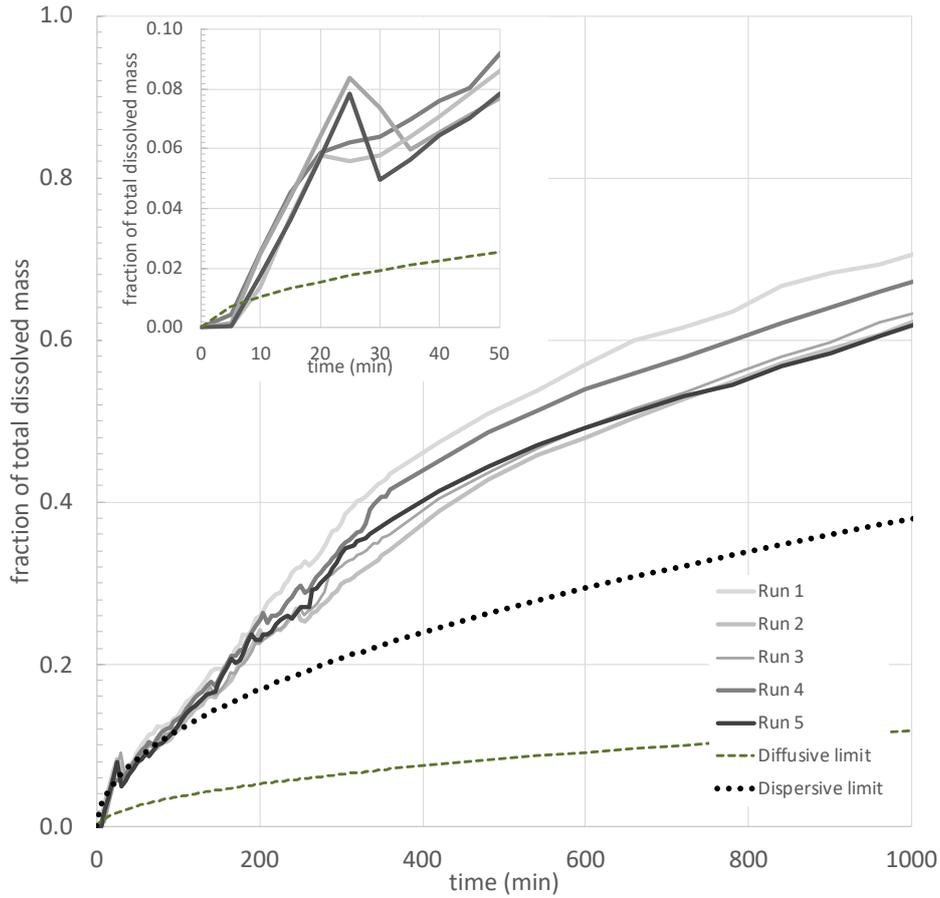

Figure 2. Summary of the dissolution of $CO_2$ into the aqueous phase for 5 repeat experiments. The diffusive limit is computed using $\mathcal{D}_{cb} = 1.9 \times 10^{-9}$ m$^2$/s and Eq. (8) from (Hassanzadeh et al., 2007). Dispersivity is found as $D_c/\mathcal{D}_c = 10$. The inset indicates that the experiments deviate from diffusive transport at times less than 10 minutes. Experimental time resolution is $\Delta t = 5$ min.

## Onset of fingering

Visually, the onset time of unstable miscible fingers between the $CO_2$ saturated and unsaturated brine is relatively rapid following the establishment of the region saturated with free-phase $CO_2$. Figure 1(a) shows potential evidence of viscous fingers after 34 min of $CO_2$ injection in the FluidFlower. By this time a distinct and widespread gas phase has formed along with a narrow region of $CO_2$-saturated brine beneath the gas zone.

Table 5. Onset times and critical wavelengths for miscible fingering measured in the FluidFlower and computed from literature results. Parameters needed for computation are taken from Tables 2 and 4. Computations consistently use $k_V$.

|  | FluidFlower | Northern Lights | Sleipner (Utsira) | In Salah (Krechba) |
|---|---|---|---|---|
| **Onset time, $t_f$** | min | year | year | year |
| experimental | < 28 |  |  |  |
| (Elenius et al., 2012) | 1.3-6.2 | 6.8-32 | 0.25-1.8 | 1500-7000 |
| (Riaz et al., 2006) | 16 | 127 | 3.2 | 44,000 |
| (Hassanzadeh et al., 2007) | 43 | 13,000 | 48 | $2.9 \times 10^6$ |
|  |  |  |  |  |
| **Critical wavelength** | cm | m | m | m |
| experimental | $4.8 \pm 0.3$ |  |  |  |
| (Elenius et al., 2012) | 0.6-2.0 | 1.6-2.0 | 0.20-0.25 | 24-29 |
| (Riaz et al., 2006) | 5.0 | 79 | 6.7 | 1,800 |
| (Hassanzadeh et al., 2007) | 2.8 | 28 | 3.5 | 410 |

A variety of analytical and numerical treatments of the instability of $CO_2$-laden brine layers and subsequent convective mixing of the brine zone beneath the gas cap are available. Notably, much of the analysis in the literature assumes the rapid accumulation of a quiescent zone of free-phase $CO_2$ atop brine followed by dissolution and gravitational instability (Ennis-King & Paterson, 2005; Hassanzadeh et al., 2007; Riaz et al., 2006). On the other hand, the results summarized in Fig. 1 support the notion that the fingering begins during active injection and the capillary fringe beneath the gas cap interacts with the $CO_2$-laden brine in the diffusive boundary layer. This interaction is predicted to reduce the time required for the onset of finger formation by up to a factor of 5 (Elenius et al., 2012).

(Elenius et al., 2012) propose that the onset time, $t_f$, for convective mixing of the brine-filled zone lies within a range incorporating time scales for horizontal and vertical flow components as

$$31 \frac{\phi^2 \mu_b^2 \mathcal{D}_{cb}}{(k \Delta \varrho_b g)^2} \leq t_f \leq 146 \frac{\phi^2 \mu_b^2 \mathcal{D}_{cb}}{(k \Delta \varrho_b g)^2} \qquad (11)$$

Due to the relatively fast advance of the gas/liquid transition zone in the FluidFlower and the absence of a remarkable period of time dominated by diffusion in Fig. 2, we substitute dispersivity for diffusion within the inequality in Eq. (11) during evaluation. Additionally, the vertical permeability is obtained as the product of $k_R$ and $k_V/k_H$ from Table 2.

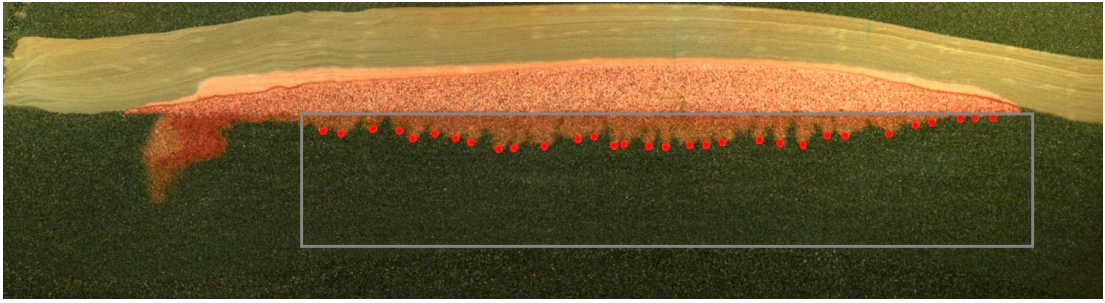

(a)

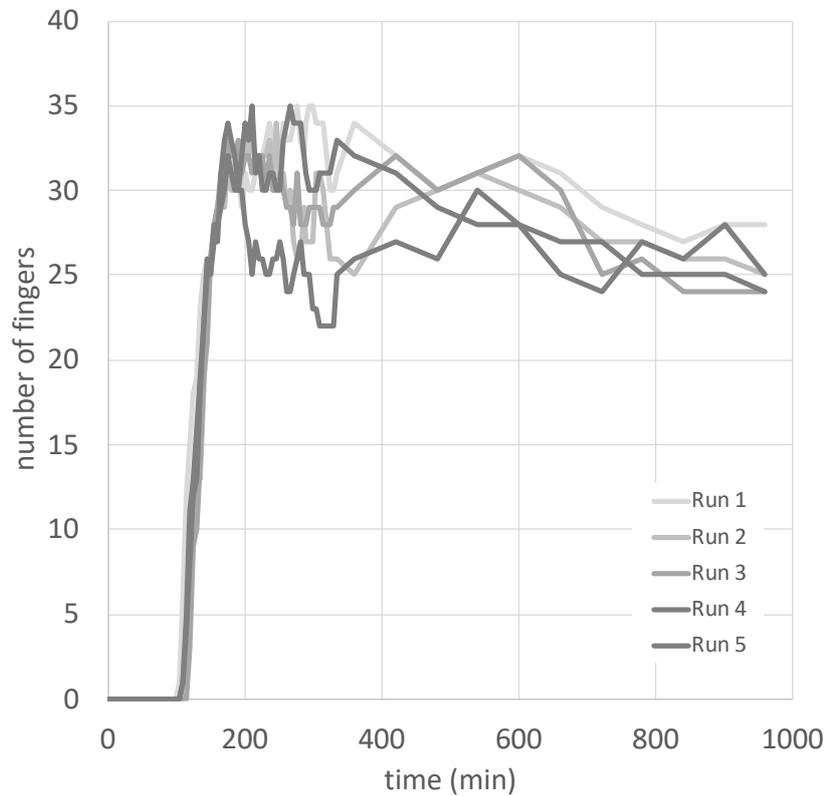

(b)

Figure 3. Analysis of fingers during experiments: (a) example labeling of fingers in a zone that is 1.5 m in length and (b) evolution of the number of fingers for 5 repeat experiments. The maximum number of fingers is found at times from roughly 160 to 250 min. During this time the average number of fingers over the 5 experiments is 31 with a standard deviation of 2.

Taking values from Table 2 and setting the difference in density to 3.5 kg/m$^3$ (Efika et al., 2016), we obtain prediction of onset time that ranges from 1 to 6 min for the FluidFlower using Eq. (11). Other predictions available from the literature are also summarized in Table 5. Experimentally, the onset time for fingering in the FluidFlower depends on the time required to build the gas-filled region in the top of the storage zone.

Onset time is found by plotting the position of fingers versus time and extrapolating the position to zero. Then, the time needed for CO$_2$ to flow from the injection point and to accumulate in the volume above the finger is subtracted to obtain the onset time. Onset times for fingering in the FluidFlower are thus found to be about 28 min. This is in order of magnitude agreement with Fig 1(a).

Experiments can also be compared to predictions of the critical wavelength, $\lambda_f$, of instabilities that grow into fingers. Similar to onset time, (Elenius et al., 2012) suggest that realistic cases for $\lambda_f$ are bounded as

$$\frac{2\pi\mu_b D_{cb}\phi}{0.086 k \Delta\varrho_b g} \leq \lambda_f \leq \frac{2\pi\mu_b D_{cb}\phi}{0.07 k \Delta\varrho_b g} \qquad (12)$$

where diffusivity has been substituted by dispersivity. With the same input as that used to evaluate Eq. (11), $\lambda_f$ ranges from 1.6 to 2.0 cm, Table 5. Experimental images are analyzed as described by (Nordbotten et al., submitted). Inspection of experimental results from the FluidFlower in Fig. 3 indicates that significant merging of fingers does not occur at relatively short times. Hence, the wavelength of macroscopic-dimension fingers at these times likely corresponds to the critical wavelength. Figure 3 indicates roughly 31 fingers below the gas cap of the storage zone in the box outlined in gray. This zone is about 1.5 m in length. Hence, the experimental wavelength of the fingers is about 4.8 ± 0.3 cm on average.

The ability of Eqs. (11) and (12) to reflect the dynamics in the FluidFlower gives us some confidence to apply them to the storage formation examples, Table 5. Generally, critical wavelengths range from 1's to 10's of m whereas onset times range from less than a year for Sleipner conditions up to 10's of thousands of years for In Salah conditions due to small $k_V$.

**End of early convective mixing**

The formation of viscous fingers marks the start of convective mixing and mixing stratifies the density gradients in the brine. This stratification of density diminishes convection and eventually leads to the reduction or gradual elimination of convection cells. Linear stability analysis is no longer applicable. Analysis of numerical simulations showed that the end of the early convective mixing period is expressed as (Hassanzadeh et al., 2007)

$$t_e = 100H^2 D_{cb}^{1/5} \left(\frac{\mu_b \phi}{\Delta \varrho_b g k H}\right)^{6/5} \tag{13}$$

where diffusivity has again been substituted with dispersivity. With the input previously used in Eqs. (11) and (12) for the FluidFlower, Eq. (13) predicts the end of the period of convective mixing driven by fingering to be 570 min in reasonable order of magnitude agreement with results in Figs. 1(c) and 2. In Fig. 2, the end of early mixing is gauged by the slope of the dissolved mass curves deviating from near constant and decreasing (Hassanzadeh et al., 2007). Note that the applicability of Eq. (13) was checked as the range of suitability is 80 < Ra < 2000. For FluidFlower conditions, Ra equals 453.

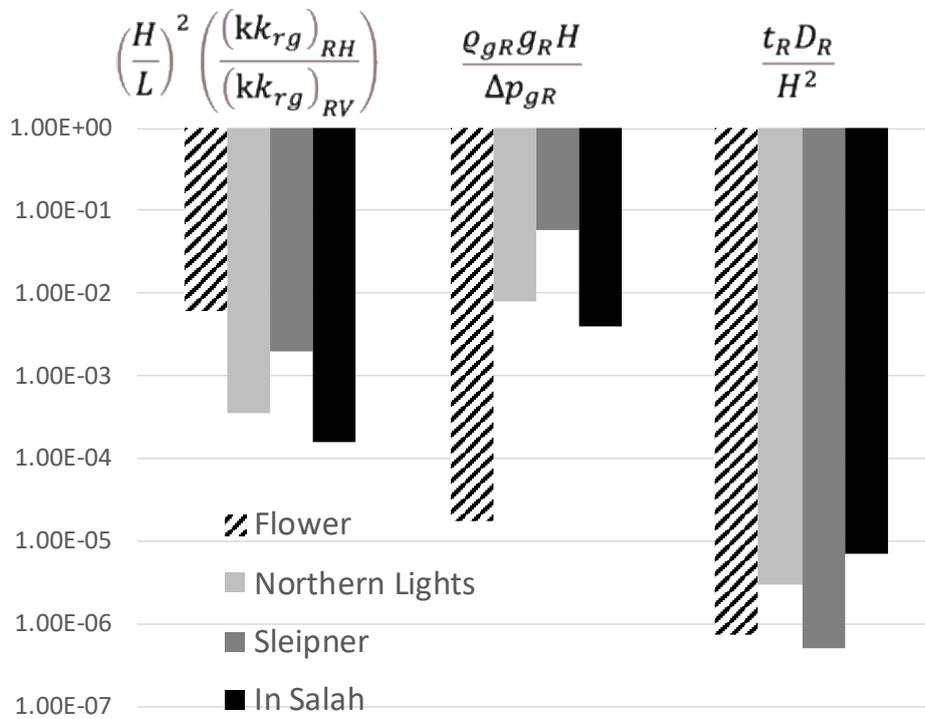

Figure 4. Summary of the magnitude of coefficients in Eq. (5) evaluated using parameters from Tables 2 and 4. Coefficients less than 1 support the importance of vertical pressure-gradient driven flow of $CO_2$ as an important transport process.

## Discussion

The physical processes considered here included convection driven by pressure gradient, convection driven by gravity, gas to brine interphase mass transfer controlled by diffusion, dispersive transport within brine, and convective mixing of dense $CO_2$-laden brine with the original aqueous phase. Visualization of experiments teaches that convection is the dominant mass transfer mechanism of both the gas and brine phases and this observation guided the scaling analysis. Figure 4 summarizes visually the magnitudes of the coefficients in Eq. (5) using characteristic parameters. It illustrates that all coefficients are at least an order of magnitude less than 1. The impact of gravity driven convection is most significant in the Northern Lights and Sleipner example field cases.

Additionally, Fig. 5 presents the difference in the relative importance of processes between the FluidFlower and the storage formations as the ratio of the scaling groups that emerged from Eq. (5) when evaluated with reference values from Tables 2 and 4. Pressure driven convection in the z direction has a ratio of 1 because the ordering process proceeded from this choice. Figure 5 shows that the x-direction pressure gradient and intraphase mass transfer are relatively greater in the FluidFlower compared to storage formations. On the other hand, gravity driven convection is relatively smaller in the FluidFlower. These predictions result in large part from the significant heights and lengths of the storage formation in comparison to the FluidFlower. For example, intraphase mass transfer in the FluidFlower is predicted to be about 100 times greater than characteristic Sleipner conditions largely due to the substantial sand thickness at Sleipner.

Convective mixing of $CO_2$-laden brine with original brine is a significant mass transfer mechanism during $CO_2$ storage (Lindeberg & Wessel-Berg, 1997; Weir et al., 1995). Application of results from linear stability analysis to FluidFlower conditions produced order of magnitude agreement for the critical wavelength of instabilities of about 4.8 cm. The critical time for the onset of instabilities, however, is predicted to differ from what is found experimentally. The physical situation in the FluidFlower does not agree with the theoretical analysis because the FluidFlower requires some time to accumulate a gas-phase region and that region subsequently grows laterally and vertically in time. Visual inspection suggests that the capillary transition zone at the base of the gas-phase region and the advancing gas/liquid interface plays a role in relatively short onset times for convective mixing. Overall, fingers in the FluidFlower establish themselves rapidly. The net effect appears to be that $CO_2$ goes into solution in brine and mixes with the original pore waters very rapidly in the FluidFlower. This aspect warrants further investigation.

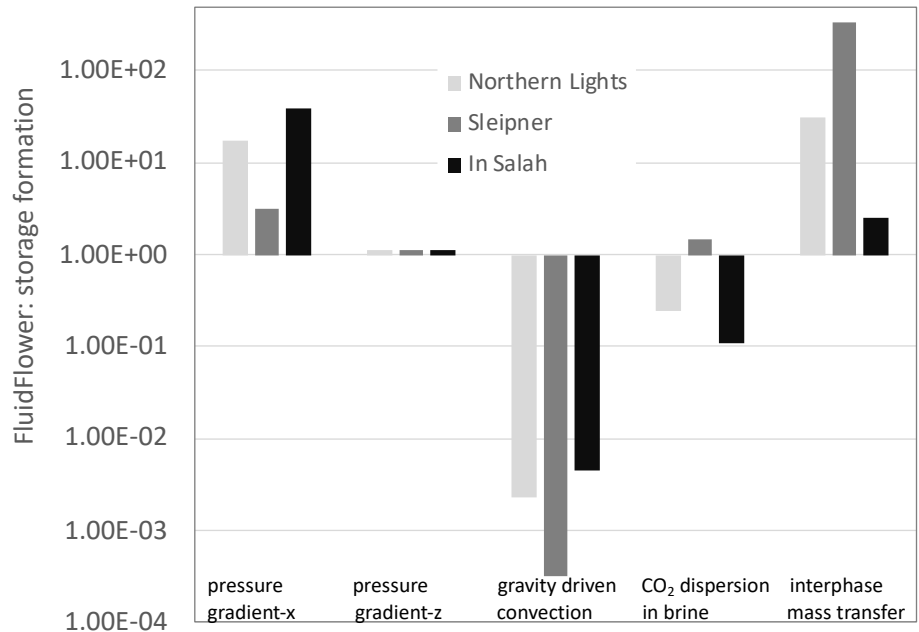

Figure 5. Ratio of scaling groups evaluated using characteristic values for FluidFlower and storage formation conditions. The ratios illustrate the relative time scales between the FluidFlower and the storage formation not the relative importance of each process. Pressure-driven convection in the z-direction has a scaling of 1 for all cases and allows estimation of storage formation time scales from FluidFlower results.

Alternate expressions to those used here for the critical onset time for fingering and the critical wavelength of fingers, Eqs. (10) and (11), were also explored. These expressions tend to produce estimates of the critical time for the onset of fingering that are substantially greater than the analysis of (Elenius et al., 2012) and somewhat closer to the experimental observations. On the other hand, estimates of the critical wavelength were all on the order of cm and in agreement with FluidFlower observations.

Interestingly, the various predictions for the onset of time of convective mixing in the case of In Salah is on the order of thousand to millions of years, but the Northern Lights and Sleipner cases were years to 10's of years. In these latter cases, convective mixing should occur during active injection into the storage formation. Accordingly, $CO_2$ goes into solution relatively quickly and convective mixing contributes to rapid downward migration of $CO_2$ during the injection period. This aspect improves storage security.

The reactions of dissolved $CO_2$ to create carbonate minerals and the dissolution of minerals were not analyzed here. The timescale of mineral trapping is 1000's to 10,000's of years in sedimentary sandstone formations that do not contain relatively abundant and soluble silicate minerals (Audigane et al., 2007; Zhang & DePaolo, 2017). Such solubilization releases calcium, magnesium, and iron species that may combine with $CO_2$. Hence, the formation of carbonate minerals is expected to contribute little to the sequestration of $CO_2$ during the period of active injection and thereafter for quite some time in the example formations used here. Formations with the potential for rapid carbon mineralization contain ultramafic igneous or metamorphic rocks, such as basalt. Incorporation of mineralization driven by low pH brine into the scaling analysis represents a significant extension due to reaction network complexity. This is left as a potential topic for future work.

## Summary and Conclusion

A general, nonequilibrium mass balance for $CO_2$ interaction with brine under immiscible conditions was analyzed to compare laboratory conditions at ambient temperature and pressures with storage formations at temperature ranging from 41 to 95 °C and pressures up to 29 MPa. The period of interest was active $CO_2$ injection operations. Comparison of the dimensionless groups developed by an ordering analysis showed that conditions in the FluidFlower and storage formations are largely dominated by convection driven by the imposed pressure gradient. The contribution of gravity to convective transport is somewhat greater in the storage formations as compared to the FluidFlower. Results indicate that the various physics seen in the FluidFlower during injection are acceptably scaled in comparison to the field given physical constraints. That is, relatively straightforward scaling of time is possible between the FluidFlower and storage formation conditions.

Significant convective mixing of $CO_2$ that has dissolved into formation brine with $CO_2$-free brine is found in the FluidFlower. The magnitude of onset time for downward migrating fingers containing $CO_2$ is only a fraction of the duration of $CO_2$ injection. Hence, the condition of quiescent fluids prior to convective mixing, as assumed in many theoretical analyses, is not met in the FluidFlower. Application of predictions for onset times to representative storage formation conditions likewise teaches that the onset time for viscous fingering is significantly less than the duration of $CO_2$ injection. The implications of this observation include that mixing of $CO_2$ with brine and the subsequent settling due to gravity may be more rapid than some prior predictions. More rapid mixing is a favorable outcome enhancing $CO_2$ storage security.

## Acknowledgement

We thank B. Benali and J.W. Both for assistance analyzing FluidFlower results. ARK acknowledges the support of the Stanford University Energy Transition Research Institute (SUETRI-A) as well as the Stanford Center for Carbon Storage (SCCS).

## Competing Interests

The authors declare no known competing interests.

## Nomenclature

| | |
|---|---|
| $a_i$ | interfacial area |
| $B$ | constant in Eq. (8) |
| $c$ | constant in Eq. (10) |
| $D$ | dispersion |
| $D$ | diffusivity |
| $\boldsymbol{g}$ | acceleration due to gravity |
| $h$ | Henry's law constant |
| $H$ | characteristic vertical dimension |
| $J_i$ | diffusive flux of component i |
| $k$ | absolute permeability |
| $K$ | mass transfer coefficient |
| $k_r$ | relative permeability |
| $L$ | characteristic horizontal length |
| $p$ | pressure |
| $q$ | injection/production rate source sink term |
| $S$ | saturation |
| *t* | time |
| *T* | absolute temperature |
| *u* | Darcy velocity |
| *v* | interstitial velocity |
| *x* | horizontal distance |
| *z* | vertical distance |

# Greek letters

| | |
|---|---|
| $\phi$ | porosity |
| $\lambda$ | wavelength |
| $\mu$ | viscosity |
| $\varrho$ | mass density |
| $\nu$ | kinematic viscosity |
| $\rho$ | molar density |
| $\tau$ | tortuosity |
| $\chi$ | mole fraction |

# subscripts and superscripts

| | |
|---|---|
| $\beta$ | phase |
| b | refers to aqueous phase |
| c | refers to $CO_2$ chemical component |
| D | dimensionless |
| f | refers to fingering |
| g | refers to $CO_2$-rich phase |
| H | refers to horizontal direction |
| i | component |
| R | reference or characteristic value |
| V | refers to vertical direction |
| w | refers to water chemical components |

## Appendix A: Mass Transfer Resistance

This appendix develops the two-film model for mass transfer resistance that is used to find overall mass transfer resistance following the ideas of (Lewis & Whitman, 1924). Figure A1 sketches bulk fluid phases with an interfacial region. The interface is marked as a black dashed line. The overall mass transfer resistance describes transfer from the g to the b phase. There are stagnant films with unequal dimensions on each side of the interface. A Henry's law relation is written as

$$\chi_{cb} = H p_{cg} \qquad (A1)$$

to describe the equilibrium solubility of $CO_2$ in the brine phase. It is assumed that only the interface is at equilibrium and so describable by Henry's law. The flux across film-1 is equal to the flux across film-2 and written as

$$J_c = K_g (p_{cg} - p_{ci}) \qquad (A2a)$$

$$J_c = K_b (\chi_{ci} - \chi_{cb}) \qquad (A2b)$$

where the subscript $i$ denotes conditions at the interface, $K_g$ and $K_b$ are mass transfer coefficients for the respective films, and it is clear that $K_g$ and $K_b$ have different units.

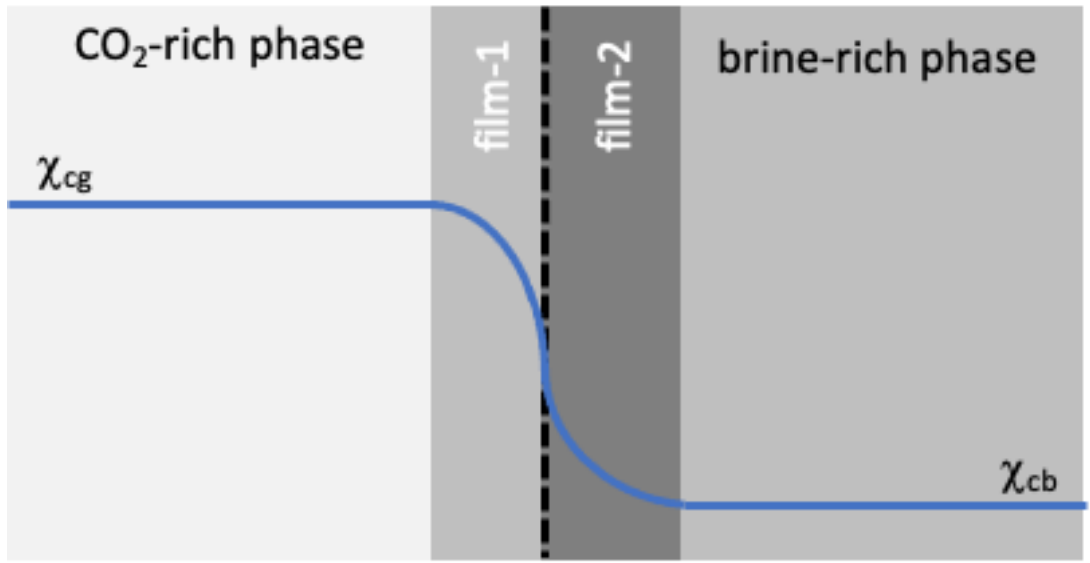

Figure A1. Schematic of the interface region in the two-film model for mass transfer. The dark dashed line indicates the interface. The shaded regions are stagnant films on either side of the interface.

To proceed, Eq. (A1) is substituted into the expression for mass transfer in Eq. (B2b) and the result solved for $p_{ci}$ to obtain

$$p_{ci} = \left(\frac{J_c}{K_b} + \chi_{cb}\right)\frac{1}{H} \tag{A3}$$

Equation (B3) is then substituted into the expression for mass transfer in the $CO_2$-rich phase in Eq. (B2a) and solved for the flux as

$$J_c = \frac{Hp_{cg} - \chi_{cb}}{\frac{H}{K_g} + \frac{1}{K_b}} \tag{A4}$$

Equation (B4) describes the flux of $CO_2$ from phase g to b and is rewritten as

$$J_c = K(Hp_{cg} - \chi_{cb}) \tag{A5}$$

where the overall mass transfer coefficient is found as

$$\frac{1}{K} = \frac{H}{K_g} + \frac{1}{K_b} \tag{A6}$$